%% file: KK1d_pPb_v10.tex
\documentclass[ALICE,manyauthors]{cernphprep}

\usepackage{amsmath}
\usepackage{bm}

\usepackage[comma,square,numbers,sort&compress]{natbib}
\usepackage{color}
\usepackage{hyperref}

\usepackage{multirow} 
\usepackage{ltxcmds}

\usepackage{placeins}
\usepackage{subfigure}

\begin{document}%

\begin{titlepage}
\PHyear{2019}
\PHnumber{054}      
\PHdate{25 March}  
%

\title{One-dimensional charged kaon femtoscopy \\ in p--Pb
collisions at {\ensuremath{\pmb{\sqrt{s_{\rm NN}}}}}{\bf ~=~5.02}~TeV}
\ShortTitle{One-dimensional charged kaon femtoscopy in p--Pb
collisions}   

\Collaboration{ALICE Collaboration\thanks{See Appendix~\ref{app:collab} for the list of collaboration members}}
\ShortAuthor{ALICE Collaboration} 

\begin{abstract}
The correlations of identical charged kaons were measured in p--Pb collisions at $\sqrt{s_{\rm NN}}=5.02$~TeV by the ALICE experiment at the LHC. The femtoscopic invariant radii and correlation strengths were extracted from one-dimensional kaon correlation functions and were compared with those obtained in pp and Pb--Pb collisions at $\sqrt{s}=7$~TeV and $\sqrt{s_{\rm NN}}=2.76$~TeV, respectively.
The presented results also complement the identical-pion femtoscopic data published by the ALICE collaboration. The extracted radii increase with increasing charged-particle multiplicity and decrease with increasing pair transverse momentum. At comparable multiplicities, the radii measured in p--Pb collisions are found to be close to those observed in pp collisions.
The obtained femtoscopic parameters are reproduced by the EPOS 3 hadronic interaction model and disfavor models with large initial size or
strong collective expansion at low multiplicities.
\end{abstract}
\end{titlepage}
\setcounter{page}{2}

\section{Introduction}
The Bose--Einstein enhancement of the production of two identical pions at low relative momenta (or, in other words, quantum statistics correlations) was first observed in the $\overline{\rm p}$p annihilation more than 50 years ago \cite{Goldhaber:1960sf}. These correlations encode information about the space--time structure of the interaction region of particles created in collisions at kinetic freeze-out (``particle-emitting source'') \cite{Kopylov:1972qw,Kopylov:1974uc,Lednicky:2003mq}. Since that time the correlation method has been developed \cite{Kopylov:1975rp,Lednicky:2005tb} and it is now known as ``correlation femtoscopy''. Femtoscopy measures the apparent width of the distribution of the relative separation of emission points, which is conventionally called the ``radius parameter''. The method was successfully applied to the measurement of the space--time characteristics of particle production processes at high energies in particle \cite{Kittel:2001zw,Alexander:2003ug} and heavy-ion collisions (see, e.g., \cite{Podgoretsky:1989bp,Lednicky:2003mq} and references therein).

Identical boson correlations, especially of identical charged pions, have been used extensively over the years to experimentally study properties of the emitting source created in various collision systems \cite{Lisa:2005dd}. Identical charged kaon femtoscopy studies were also carried out, for example, in Au--Au collisions at $\sqrt{s_{\rm NN}}=200$~GeV by the STAR \cite{Adamczyk:2013wqm} and PHENIX \cite{Afanasiev:2009ii} collaborations and in pp collisions at $\sqrt s=~7$~TeV and Pb--Pb collisions at $\sqrt{s_{\rm NN}}=2.76$~TeV by the ALICE collaboration \cite{Abelev:2012sq,Adam:2015vja}.

The study of femtoscopic correlations in asymmetric collision systems is particularly interesting because it provides a bridge between small (pp) and large (A--A) collision systems, and may lead to additional constraints on model scenarios which were successfully used to describe pp and A--A collisions. The A--A femtoscopy results are interpreted within the hydrodynamic framework as a signature of collective radial flow \cite{Lisa:2005dd,Hirono:2014dda,Soltz:2012rk,Karpenko:2012yf}. Attempts to describe the pp data in the same framework have not been successful so far and it is speculated that additional effects related to the uncertainty principle may play a role in such small systems \cite{Shapoval:2013jca}. The results obtained in asymmetric collisions are difficult to interpret unambiguously. For instance, the femtoscopic study of the data obtained at RHIC for d--Au collisions \cite{Adare:2014vri,Bozek:2014fxa} suggest that a hydrodynamic evolution may be present in such a system, while at the LHC the ALICE three-pion \cite{Abelev:2014pja} and three-dimensional two-pion \cite{Adam:2015pya} analyses in p--Pb collisions at $\sqrt{s_{\mathrm{NN}}}=5.02$~TeV demonstrate the more important role of the initial state shape and size of the created system.

The excellent particle identification capabilities of the ALICE detector \cite{Aamodt:2008zz} and the data sample collected in p--Pb collisions at $\sqrt{s_{\mathrm{NN}}}=5.02$~TeV in 2013 allow one to perform the K$^{\pm}$K$^{\pm}$ femtoscopic analysis.
Kaons are a convenient tool to study Bose--Einstein correlations because they are less influenced by resonance decays than pions and therefore more effectively probe femtoscopic correlations of directly-produced particles. The comparison of kaon and pion correlation radii \cite{Adam:2015pya,Abelev:2014pja,Adam:2015vja} as a function of pair transverse momentum $k_{\rm T}=|{\bf p}_{\mathrm{T,1}} + {\bf p}_{\mathrm{T,2}}|/2$ or transverse mass $m_{\rm T}= \sqrt{k_\mathrm{T}^2+m^{2}}$, where ${\bf p}_{\mathrm{T,1}}$ (${\bf p}_{\mathrm{T,2}}$) is the transverse momentum of the first (second) particle and $m$ is the kaon or pion mass, allows one to understand the collective dynamics (collective flow) of the source created in high-energy collisions. In particular, in the system created by colliding heavy ions, the decrease of the correlation radii with increasing $k_{\mathrm T}$ ($m_{\mathrm T}$) is usually considered as a manifestation of the strong collective expansion of the matter created in such collisions. If the dependence of the interferometry radii on pair momentum in p--Pb collisions followed the trends seen in heavy-ion collisions, it would be an indication of collectivity or the creation of a hot and dense system expanding hydrodynamically \cite{Bozek:2013df,Pierog:2013ria,Werner:2013tya}.
In addition, comparing kaon femtoscopic results in pp, p--A, and A--A collision systems can provide experimental constraints on the validity of hydrodynamic \cite{Bozek:2013df,Werner:2013tya} and/or color glass condensate \cite{Dusling:2013qoz,Bzdak:2013zma} approaches proposed for the interpretation of the p--Pb data.
In this work, the kaon femtoscopic radii in p--Pb collisions at $\sqrt{s_{\mathrm{NN}}}=5.02$~TeV are shown as a function of $k_{\mathrm T}$ and multiplicity, and are compared with those in pp and Pb--Pb collisions at $\sqrt{s}=7$~TeV \cite{Abelev:2012sq} and $\sqrt{s_{\mathrm{NN}}}=2.76$~TeV \cite{Adam:2015vja}, respectively.
The presented data are also compared with the EPOS 3.111 model \cite{Werner:2013tya}, an event generator based on a (3+1)D viscous hydrodynamical evolution starting from flux tube initial conditions, which are generated in the Gribov-Regge multiple scattering framework. The approach contains a full viscous hydrodynamical simulation and a more sophisticated treatment of nonlinear effects in the parton evolution by considering individual (per Pomeron) saturation scales than in previous EPOS versions \cite{Werner:2010aa,Pierog:2013ria}.
There are also changes in the core--corona procedure \cite{Aichelin:2008mi} crucial in proton--nucleus collisions, so that the initial energy of the flux tubes is separated into a part which constitutes the initial conditions for hydrodynamic expansion (core) and the particles which leave the ``matter'' (corona). This model reasonably reproduces multiplicity distributions, transverse momentum spectra, and flow results, and it gives the best description of kaon spectra \cite{Drescher:2000ha,Werner:2010aa,Pierog:2013ria,Werner:2015vza,Acharya:2018egz}.

The paper is organized as follows: Section 2 shortly describes the ALICE experimental setup and charged kaon selection criteria used in the presented work. In Section 3, the femtoscopic correlation function analysis is described in detail and the sources of systematic uncertainties are discussed. The extracted radii and correlation strengths are shown and compared with model predictions in Section 4. The obtained results are summarized in Section 5.

\section{Description of the experiment and data selection}

\subsection{Experiment}
The ALICE detector and its performance in the LHC Run 1 (2009 -- 2013) are described in \cite{Aamodt:2008zz} and \cite{Abelev:2014ffa}, respectively. About 55$\times$10$^6$ p--Pb collision events collected in 2013 at a center-of-mass energy per nucleon--nucleon pair of $\sqrt{s_{\mathrm{NN}}}=5.02$~TeV were analyzed in this work. Given the energies of the colliding p and Pb beams, the nucleon--nucleon center-of-mass system is shifted with respect to the ALICE laboratory system by 0.465 units of rapidity in the direction of the proton beam. Throughout this paper $\eta$ represents the pseudorapidity measured in the laboratory frame.

The analyzed events were classified according to their multiplicity \cite{Abelev:2013haa,Adam:2014qja} using the measured energy deposition in the V0 detectors \cite{Abbas:2013taa}, which consist of two arrays of scintillators located along the beamline installed on each side of the interaction point and covering $2.8<\eta<5.1$ (V0A, located on the Pb-remnant side) and $-3.7 < \eta < -1.7$ (V0C) \cite{Abelev:2013vea}. Charged kaons were reconstructed with the central barrel detectors placed inside a solenoidal magnet providing a 0.5~T field parallel to the beam direction, namely the Time Projection Chamber (TPC) \cite{Alme:2010ke} and the Inner Tracking System (ITS) \cite{Aamodt:2008zz}. The primary vertex was obtained from the ITS. Its position along the beam direction (the ``$z$-position'') was required to be within $\pm$10~cm of the center of the ALICE detector to ensure uniform tracking performance. The TPC was used to reconstruct tracks and their momenta. The TPC is divided by the central electrode into two halves, each of which is composed of 18 sectors (covering the full azimuthal angle) with 159 pad rows placed radially in each sector. A track signal in the TPC consists of space points (clusters), each of which is reconstructed in one of the pad rows.
The TPC covers an acceptance of $|\eta| < 0.8$ for tracks which reach the outer radius of the detector and $|\eta| < 1.5$ for shorter tracks.
The parameters of the track were determined by performing a Kalman fit to a set of clusters with an additional constraint that the track passes through the primary vertex. The quality of the fit is required to have $\chi^2/{\rm NDF}$ less than 2. The transverse momentum of each track was determined from its curvature in the uniform magnetic field.
The track selection criteria based on the quality of the track reconstruction fit and the number of detected tracking points in the TPC \cite{Abelev:2014ffa,Akindinov:2013tea} were used to ensure that only well-reconstructed tracks were considered in the analysis.

Particle identification (PID) for reconstructed tracks was carried out using the TPC together with the Time-of-Flight (TOF) \cite{Akindinov:2013tea} detector. The TOF is a cylindrical detector 
consisting of 18 azimuthal sectors divided into five modules along
the beam axis at a radius $r\simeq380$~cm. The active elements are multigap resistive plate chambers. For TPC PID, a parametrized Bethe-Bloch formula was used to calculate the specific energy loss ${\rm d}E/{\rm d}x$ in the detector expected for a particle with a given mass, charge, and momentum. For PID based on TOF information, the particle mass was used to calculate the expected time-of-flight as a function of track length and momentum. For each PID method, the signal for each reconstructed particle is compared with the one expected for a kaon taking into account the detector resolution. The allowed deviations ($n_\sigma$) depend on the momentum of the particle \cite{Adam:2015vja,Abelev:2014ffa,Akindinov:2013tea}.

\subsection{Charged kaon selection}\label{ChargedKaonSelection}
Track reconstruction for the charged kaon analysis was performed using the signals in the TPC. To ensure a good momentum resolution, each track was required to be composed of at least 70 out of the 159 TPC clusters.
Tracks were selected based on their distance of closest approach (DCA) to the primary vertex, which was required to be less than 2.4~cm in the transverse plane and less than 3.0~cm in the longitudinal direction. The kinematic range for kaons selected in this analysis is $0.14<p_{\rm T}<1.5$~GeV/$c$ and $|\eta|<0.8$.
Charged tracks with momentum $p < 0.5$~GeV/$c$ were identified as kaons if they satisfied the requirement $n_{\sigma,{\rm TPC}}<2$ in the TPC. Tracks with $p > 0.5$~GeV/$c$ were required to match to a signal in the TOF, and satisfy $n_{\sigma,{\rm TPC}}<3$ as well as the following momentum-dependent $n_{\sigma}$ selection: $n_{\sigma,{\rm TOF}}<2$ for $0.5<p<0.8$~GeV/$c$, $n_{\sigma,{\rm TOF}}<1.5$ for $0.8 < p< 1.0$~GeV/$c$ and $n_{\sigma,{\rm TOF}}<1$ for $1.0 < p< 1.5$~GeV/$c$.
All selection criteria are listed in Tab.~\ref{tab:KKcuts}.
\begin{table}
\centering
\caption{Charged kaon selection criteria.}
\begin{tabular}{l|l}
  \hline
    $p_{\rm T}$  & $0.14<p_{\rm T}<1.5$ GeV/$c$ \\ \hline
    $|\eta|$ & $< 0.8$ \\ \hline
    $\rm DCA_{\rm transverse}$ to primary vertex & $< 2.4$ cm \\ \hline
    $\rm DCA_{\rm longitudinal}$ to primary vertex & $< 3.0$ cm \\ \hline
    $n_{\sigma,\rm TPC}$ (for $p < 0.5$~GeV/$c$) & $< 2$ \\ \hline
    $n_{\sigma,\rm TPC}$ (for $p > 0.5$~GeV/$c$) & $< 3$ \\ \hline
    $n_{\sigma,\rm TOF}$ (for $0.5 < p < 0.8$ GeV/$c$) & $< 2$ \\ \hline
    $n_{\sigma,\rm TOF}$ (for $0.8 < p < 1.0$ GeV/$c$) & $< 1.5$ \\ \hline
    $n_{\sigma,\rm TOF}$ (for $1.0 < p < 1.5$ GeV/$c$) & $< 1.0$ \\ \hline
    Number of track points in TPC  & $\geq$70 \\ \hline
\end{tabular}
\label{tab:KKcuts}
\end{table}

\begin{figure}[h]
\centering
\subfigure{\label{fig:single_purity}\includegraphics[width=0.49\linewidth]{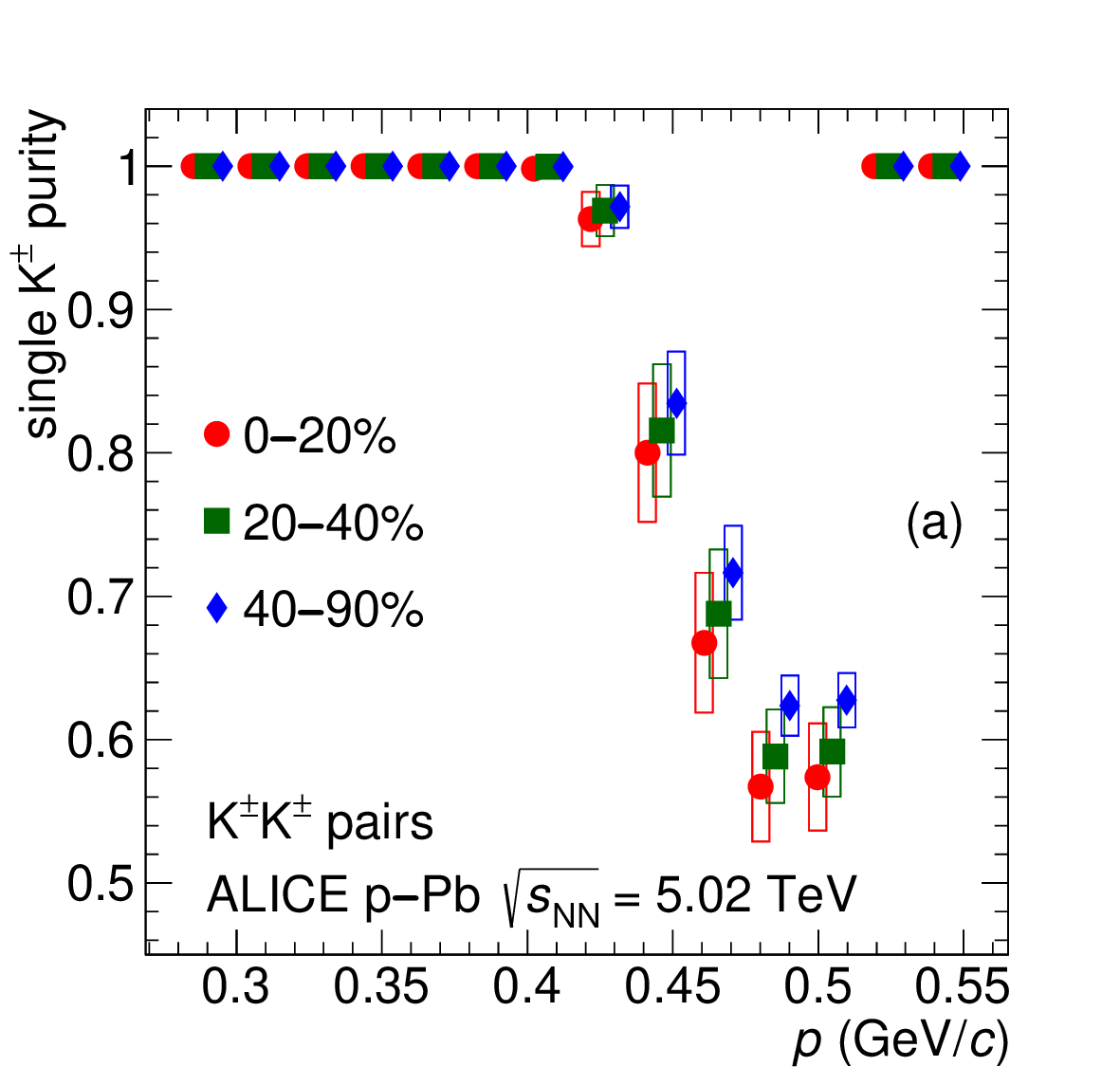}}
\subfigure{\label{fig:pair_purity}\includegraphics[width=0.49\linewidth]{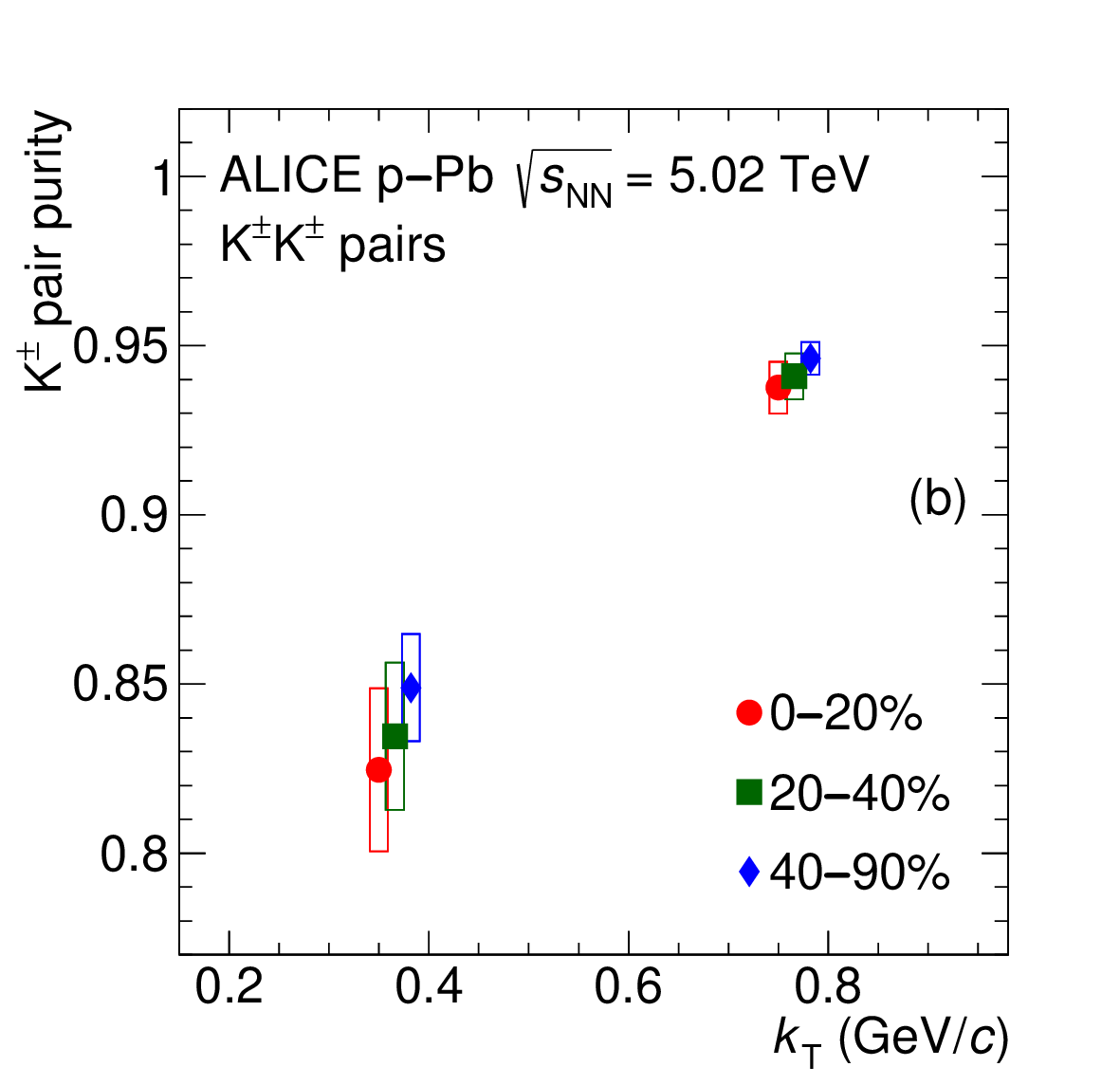}}
\caption{(color online) Single (a) and pair (b) K$^\pm$ purities for different event multiplicity classes. The systematic uncertainties associated with the purity correction are shown as boxes. Statistical uncertainties are negligible. The momentum $p$ ($k_{\rm T}$) values for lower multiplicity classes (blue and green symbols) are slightly offset for clarity.}\label{fig:purity}
\end{figure}

The estimation of purity for $p<0.5$~GeV/$c$ was performed by parametrizing the TPC d$E$/d$x$ distribution of the experimental data in momentum slices and computing the fraction of particle species that could mistakenly contribute to the kaon signal \cite{Abelev:2014ffa}. The use of momentum-dependent values for $n_{\sigma,{\rm TPC}}$ and $n_{\sigma,{\rm TOF}}$ was the result of studies to obtain the best kaon purity, defined as the fraction of accepted kaon tracks that correspond to true kaons, while retaining a decent efficiency of the PID.
The dominant contamination for charged kaons comes from e$^\pm$ in the momentum range $0.4<p<0.5$~GeV/$c$.
The parameters of the function which fits the TPC distribution in momentum slices depend on the fit interval and can be a source of the systematic uncertainty associated with the single purity.
The purity for $p>0.5$~GeV/$c$, where the TOF information was employed, was studied with DPMJET \cite{Roesler:2000he} simulations using GEANT \cite{Brun:1994aa} to model particle transport through the detector. Based on the results of this study, the $n_{\sigma,{\rm TOF}}$ values were chosen to provide a charged kaon purity greater than 99\%.
The momentum dependence of the single kaon purity in the region of maximal contamination is shown in Fig.~\ref{fig:single_purity}. The pair purity is calculated as the product of two single-particle purities for pairs with $q_{\rm inv}<0.25$~GeV/$c$, where the momenta are taken from the experimentally determined distribution. The obtained K$^\pm$ pair purity is shown in Fig.~\ref{fig:pair_purity} as a function of $k_{\rm T}$.
It can be seen from the figure that, despite the lower purity for single kaons in the range $0.45 < p < 0.5$~GeV/$c$, the pair purity remains high in the wide $k_{\mathrm T}$ bins used in the analysis due to the effects of averaging over low-purity $0.45<p<0.5$~GeV/$c$ and high-purity $p<0.45$~GeV/$c$ or $p>0.5$~GeV/$c$ bins in the full single-kaon momentum range. The systematic uncertainties of the single purity values lead, in turn, to systematic uncertainties of the obtained pair purity.

The analysis was performed in three event multiplicity classes \cite{Abelev:2013qoq,Abelev:2013haa,Adam:2014qja}: 0--20\%, 20--40\% and 40--90\% and two pair transverse momentum $k_{\mathrm T}$ bins: (0.2--0.5) and (0.5--1.0)~GeV/$c$. The multiplicity was determined based on the sum of the signal amplitudes of V0A and V0C detectors, commonly referred to as V0M. Table \ref{tab:KKmultiplicities} shows the corresponding mean charged-particle multiplicity densities $\langle {\rm d}N_{\rm ch}/{\rm d}\eta\rangle$ averaged over $|\eta|<0.5$ using the method presented in \cite{Abelev:2013haa}. The $\langle {\rm d}N_{\rm ch}/{\rm d}\eta\rangle$ values were not corrected for trigger and vertex-reconstruction inefficiency, which is about 4\% for non-single diffractive events \cite{ALICE:2012xs}.
At least one particle in the event had to be reconstructed and identified as a charged kaon. The correlation signal was constructed from events having at least two identical charged kaons.
Events with a single kaon were included in the event mixing procedure to determine the reference distribution.
\begin{table}[ht]
\centering
\caption{V0M event classes and their corresponding $\langle {\rm d}N_{\rm ch}/{\rm d}\eta\rangle$ \cite{Abelev:2013haa}. The given uncertainties are systematic only since the statistical uncertainties are negligible.}
\begin{tabular}{c|c}
  \hline
    Event class  &  $\langle {\rm d}N_{\rm ch}/{\rm d}\eta\rangle$, $|\eta|<0.5$\\ \hline
    0--20\%       &  47.3$\pm$0.7\\ \hline
    20--40\%      &  24.3$\pm$0.7\\ \hline
    40--90\%      &  17.3$\pm$1.5\\ \hline
\end{tabular}
\label{tab:KKmultiplicities}
\end{table}

The femtoscopic correlation functions (CF) of identical particles are sensitive to two-track reconstruction effects because the considered particles are close in momentum and have close trajectories. Two kinds of two-track effects were investigated. Track ``splitting'' occurs when one track is mistakenly reconstructed as two. Track ``merging'' is the effect when two different tracks are reconstructed as one. To remove these effects, pairs with relative pseudorapidity $|\Delta\eta|<0.02$ and relative azimuthal angle $|\Delta\varphi^*|<0.045$ were rejected. The modified azimuthal angle $\varphi^*$ takes into account the bending of the tracks inside the  magnetic field and was calculated at a radial distance of 1.2~m \cite{Adam:2015vna}. 

\section{\label{CFan}Analysis technique}
The correlation function of two particles with momenta ${\bf p_1}$ and ${\bf p_2}$ is defined as a ratio
\begin{equation}
C({\bf p}_1,{\bf p}_2)=\frac{A({\bf p}_1,{\bf p}_2)}{B({\bf p}_1,{\bf p}_2)}
\end{equation}
of the two-particle distribution in the given event $A({\bf p}_1,{\bf p}_2)$ to the reference distribution $B({\bf p}_1,{\bf p}_2)$ \cite{Kopylov:1974th}. The reference distribution is formed by mixing events containing at least one charged kaon, where each event is mixed with five other events which have similar $z$ position of the primary vertex and similar multiplicity \cite{Lisa:2005dd}.
The mixed particles come from events for which the vertex positions in the beam direction agree within 2~cm and the multiplicities do not differ by more than 1/4 of the width of the given multiplicity class.
The correlation function is measured as a function of the invariant pair relative momentum
$q_\mathrm{inv}= \sqrt{|{\bf q}|^{2} - q_{0}^{2}} $, where $q_0=E_1-E_2$ and ${\bf q}={\bf p_1}-{\bf p_2}$ are determined by the energy components $E_1$, $E_2$ and momenta ${\bf p_1}$, ${\bf p_2}$ of the particles, respectively.
The correlation function is normalized to unity such that $C\rightarrow1$ in the absence of a correlation signal.

The obtained correlation function $C_{\rm raw}$ was also corrected for purity before the fit \cite{Adam:2015vja,Acharya:2017qtq} according to
\begin{equation}
C_{\rm corrected} = (C_{\rm raw}-1+P)/P,
\label{eq:PurCorr}
\end{equation}
where the pair purity $P$ is taken from Fig.~\ref{fig:pair_purity}.

\subsection{Correlation function parametrization}
The CFs can be parametrized by various formulae depending on the origin of correlations between the considered particles. The pairwise interactions between K$^\pm$K$^\pm$ that form the basis for femtoscopy are quantum statistics and the Coulomb interaction. Strong final-state interactions between kaons are negligible \cite{Beane:2007uh}.
Assuming a Gaussian distribution of a particle source
in the pair rest frame, the fit of the kaon CF is performed using the Bowler--Sinyukov formula \cite{Sinyukov:1998fc,Bowler:1986ta}
\begin{equation}
C(q_\mathrm{inv})=N\left[1 -\lambda +\lambda K(r,q_\mathrm{inv})\left( 1+
\exp{\left(-R_\mathrm{inv}^{2} q^{2}_\mathrm{inv}\right)}\right)\right]\,D(q_\mathrm{inv}).
\label{eq:BS_CF}
\end{equation}
The factor $K(r,q_\mathrm{inv})$ describes the Coulomb interaction with a radius $r$, $D(q_\mathrm{inv})$ parametrizes the baseline including all non-femtoscopic
effects, for instance resonance decays, and $N$ is a normalization coefficient. The Coulomb interaction is determined as
\begin{eqnarray}
K(r,q_\mathrm{inv})=\frac{C_{\rm QS+Coulomb}}{C_{\rm QS}},\label{Coulomb}
\end{eqnarray}
where $C_{\rm QS}$ is a theoretical CF calculated with pure quantum statistical (QS) weights (wave function squared) and $C_{\rm QS+Coulomb}$ corresponds to quantum statistical plus Coulomb weights \cite{Sinyukov:1998fc,Abelev:2013pqa}.
The parameters $R_\mathrm{inv}$ and $\lambda$ describe the size of the source and the correlation strength, respectively.

\subsection{Fitting procedure}\label{FittingProcedure}
\begin{figure}
\begin{center}
\includegraphics[width=0.88\textwidth]{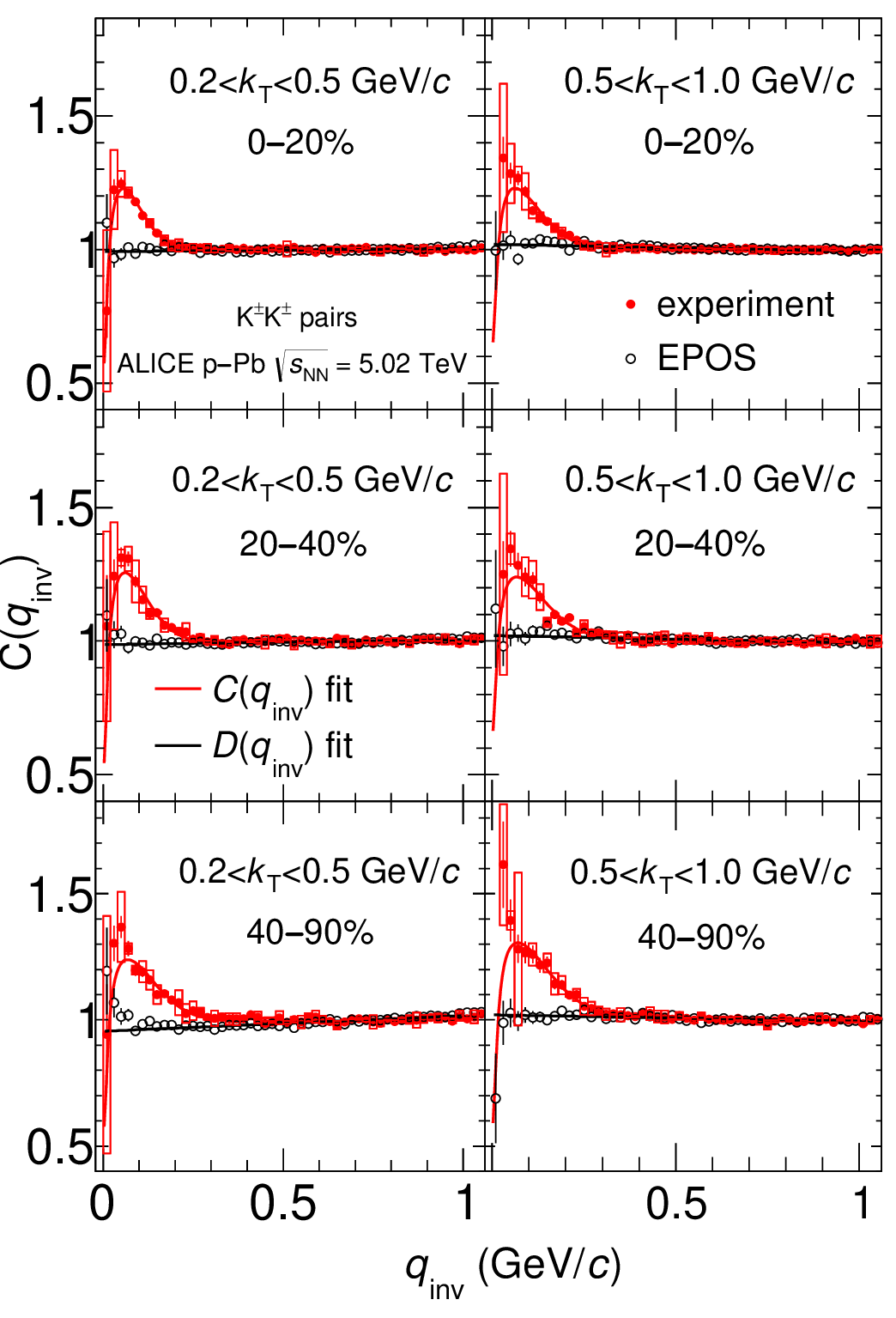}
\caption{
(color online) K$^\pm$K$^\pm$ experimental correlation functions corrected for purity according to Eq.~(\ref{eq:PurCorr}) (red points) and EPOS 3 model baselines \cite{Werner:2013tya} (black points) versus pair relative invariant momentum $q_\mathrm{inv}$. The CFs are presented in three event multiplicity classes: 0--20\%, 20--40\% and 40--90\% and two pair transverse momentum $k_{\mathrm T}$ bins: (0.2--0.5) and (0.5--1.0)~GeV/$c$. The black line shows the fit of EPOS 3 by a first-order polynomial for $0<q_\mathrm{inv}<1.0$~GeV/$c$. The red line shows the subsequent fit of the CF up to $q_\mathrm{inv}<$0.5~GeV/$c$ by Eq.~(\ref{eq:BS_CF}). The CFs are normalized to unity in the range $0.5<q_\mathrm{inv}<1.0$~GeV/$c$. Statistical (lines) and systematic uncertainties (boxes) are shown.
}
\label{fig:2017-Jun-05-CFs_fit_exp_epos}
\end{center}
\end{figure}

The parameters $R_\mathrm{inv}$ and $\lambda$ can be extracted using Eq.~(\ref{eq:BS_CF}) with various assumptions to handle the non-femtoscopic baseline $D$ from background effects outside the femtoscopic peak region.
There are various methods to deal with the baseline. The simplest way is to assume that it is flat, $D(q_\mathrm{inv})=1$, which can be reasonable in cases where non-femtoscopic effects are negligible. As can be seen in Fig.~\ref{fig:2017-Jun-05-CFs_fit_exp_epos}, the baseline of the obtained experimental functions is not flat. Therefore, it would be more reasonable to describe it, for instance, by a first-order polynomial function $D(q_\mathrm{inv})=N(1+aq_\mathrm{inv})$ which reproduces this baseline slope. It interpolates the baseline behavior at high $q_\mathrm{inv}$ taking into account all non-femtoscopic effects which make it non-flat. Then being extrapolated to low $q_\mathrm{inv}$, it is supposed to imitate the existing non-femtoscopic effects.
The most natural way to describe the baseline is to use Monte Carlo (MC) models where events are generated from physical considerations and contain all but QS and Coulomb effects. A suitable MC model has to reasonably describe the baseline at high $q_\mathrm{inv}$, where non-femtoscopic effects are significant, and also contain non-femtoscopic effects at low $q_\mathrm{inv}$.
The EPOS 3 \cite{Werner:2013tya} model without QS and Coulomb interaction effects included was used to describe the baseline $D(q_\mathrm{inv})$. As seen from Fig.~\ref{fig:2017-Jun-05-CFs_fit_exp_epos}, EPOS 3 describes the experimental CF outside the correlation peak. The extracted parameter values depend on the fit range, which should be chosen taking into account the characteristic width of the femtoscopic effect observed. In this analysis, the EPOS 3 baseline was fit with a first-order polynomial in $0<q_{\rm inv}<1.0$~GeV/$c$ (to flatten statistical uncertainties) and then the experimental CF was fit with Eq.~(\ref{eq:BS_CF}) in the range $0<q_{\rm inv}<0.5$~GeV/$c$.
The Coulomb interaction radius was set to $r=1.5$~fm, which is on average close to the extracted radii values.

\subsection{Systematic uncertainties}\label{SystematicUncertainties}
The effects of various sources of systematic uncertainty on the extracted parameters were studied as functions of multiplicity and $k_{\rm T}$. The systematic uncertainties were estimated by varying the selection criteria used for the events, particles, and pairs (with variation limits up to $\pm$20$\%$).
The influence of the fit range was estimated by variation of the $q_{\rm inv}$ upper limit by $\pm$40$\%$.
Another source of systematic uncertainty is the misidentification of particles and the associated purity correction. A $\pm$10\% variation of the parameters (Sec. \ref{ChargedKaonSelection}) used for the purity correction estimation was performed. To reduce the electron contamination, the PID criteria were tightened, in particular by extending the momentum range where the TOF signal was used and the energy-loss measurement was required to be consistent with the kaon hypothesis within $n_{\sigma,{\rm TPC}}<1$.

There is also an uncertainty associated with the choice of the radius of the Coulomb interaction. It was set to 1.5~fm as a result of averaging of the three radii values that were extracted from the respective multiplicity bins and varied by $\pm0.5$~fm. The relative difference was taken as a systematic uncertainty. Uncertainties associated with momentum resolution were estimated using a MC simulation with the DPMJET 3.05 \cite{Roesler:2000he} model. The effect is limited to low pair relative momentum, where it smears the correlation function and is especially pronounced for narrow femtoscopic peaks. In p--Pb collisions the $q_{\rm inv}$ region of the femtoscopic effect is one order of magnitude wider than the region affected by this inefficiency and, consequently, the corresponding uncertainty is minor.

As was explained in Sec.~\ref{FittingProcedure}, the fitting procedure requires knowledge of the non-femtoscopic background shape and magnitude.
In this analysis, the EPOS 3 model was used for this purpose.
The systematic uncertainty associated with the baseline was estimated using an alternative MC model, DPMJET, as well as the two methods based on the use of polynomials described in Sec.~\ref{FittingProcedure}.

Table~\ref{tab:systerrKch} presents the uncertainty range for all considered sources of systematic uncertainty, where the minimum (maximum) was chosen from all available values in all multiplicity and $k_{\rm T}$ bins.
For each source and each multiplicity and $k_{\rm T}$ bin, the maximum deviation from the parameters obtained with the optimal data selection criteria and fitting methods was taken and applied symmetrically as the uncertainty. The limited data sample for p--Pb collisions leads to quite high statistical uncertainty values and most of the systematic uncertainty contributions were found to be much smaller than the quadratic difference of the statistical uncertainties. Therefore, the systematic uncertainty values were added in quadrature, considering only those whose statistical significance level exceeded 50\% \cite{Barlow:2002yb}. As can be seen from Table~\ref{tab:systerrKch}, the main sources of systematic uncertainty on the extracted parameters are the pair selection criteria, the influence of the fit range, the radius of the Coulomb interaction and the baseline description. All of them contribute to the uncertainty associated with the radii. The extracted correlation strengths have higher statistical uncertainties than the radii and, consequently, for them the pair selection criteria is the only source of systematic uncertainty which exceeds the statistical significance level chosen in this analysis.

\begin{table}
 \centering
\caption{Minimum and maximum uncertainty values for various sources of systematic uncertainty (in percent), the punctuation `--' means that the contribution from the given source is negligible.
Note that each value is the minimum--maximum uncertainty from a specific source, but can pertain to different multiplicity or $k_{\rm T}$ bins.
Thus, the maximum total uncertainties are smaller than (or equal to) the sum of the maximum individual uncertainties shown in this table. Systematic uncertainties whose statistical significance level exceeds 50\% were included in the total systematic uncertainty value.}
 \begin{tabular}{l|c|c}
 \hline
 & $R_{\rm inv}$ (\%) & $\lambda$ (\%) \\ \hline
 {Single particle selection} & 0--1.5 & 0--3.2 \\ \hline
 {PID and purity} & -- &  0--0.6 \\ \hline
 {Pair selection} & {0--3} & {0--6} \\ \hline
 {Baseline} & {1--4.8} & 0.2--4.1 \\ \hline
 {Fit range} & {0.6--7} & 0.5--5.7 \\ \hline
 {Coulomb function} & {0--2.3} & 1.8--3.8 \\ \hline
 {Momentum resolution} & 0--1 & 0--1\\ \hline
 \end{tabular}
\label{tab:systerrKch}
\end{table}
\FloatBarrier

\section{Results and discussion}

\begin{figure}
\centering
\subfigure{\label{fig:RL_sys_EPOSA}\includegraphics[width=0.49\linewidth]{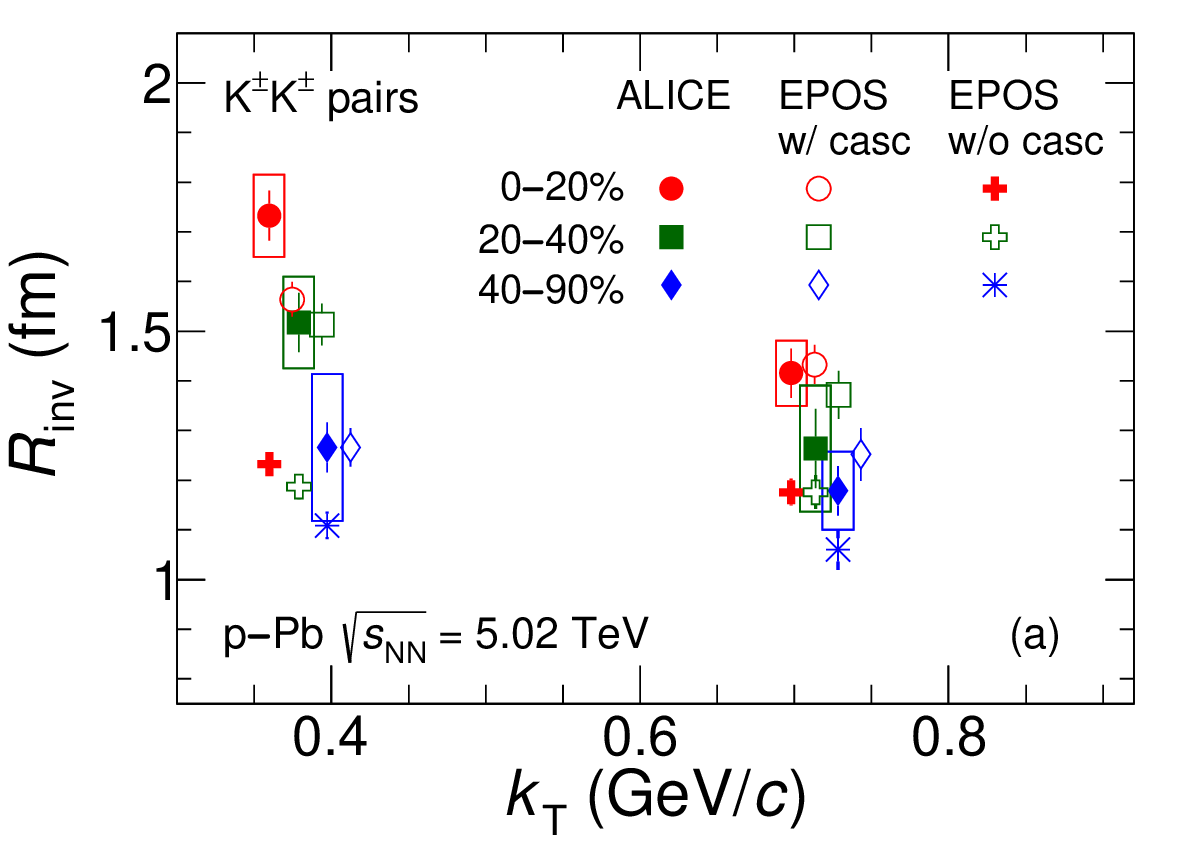}}
\subfigure{\label{fig:RL_sys_EPOSB}\includegraphics[width=0.49\linewidth]{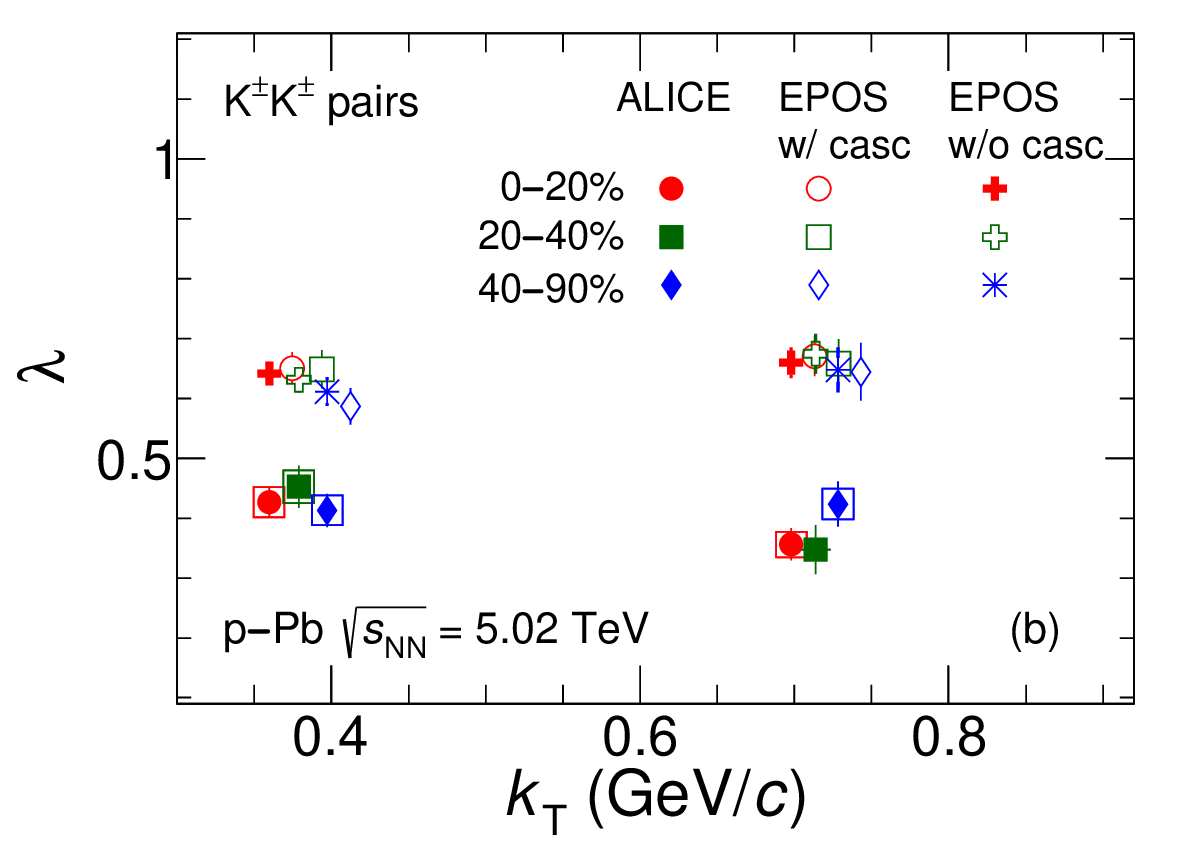}}
\caption{(color online) Experimental K$^\pm$K$^\pm$ invariant radii
$R_\mathrm{inv}$ (a) and correlation strengths $\lambda$ (b) shown versus pair transverse momentum $k_{\mathrm T}$ for three multiplicity classes and compared with the EPOS 3 model predictions with and without the hadronic cascade phase.
Statistical (lines) and systematic uncertainties (boxes) are shown. The points for lower multiplicity classes (blue and green symbols) are slightly offset in the $x$ direction for clarity.} \label{fig:RL_sys_EPOS}
\end{figure}

The extracted $R_\mathrm{inv}$ and $\lambda$ parameters are depicted in Figs.~\ref{fig:RL_sys_EPOSA} and \ref{fig:RL_sys_EPOSB}, respectively.
Statistical and systematic uncertainties as described in Sec.~\ref{SystematicUncertainties} are shown for all results.
Figure \ref{fig:RL_sys_EPOS} also shows comparisons with the EPOS 3 model (with femtoscopic effects included \cite{Werner:2010aa,Werner:2011fd,Werner:2010ny}) for the same collision system and energy in the same multiplicity and $k_{\rm T}$ bins as the experimental data. Two cases are considered, one with and another one without the hadronic cascade (UrQMD) phase \cite{Knospe:2015nva}. The EPOS 3 calculations for the radii without the cascade exhibit practically no $k_{\mathrm{T}}$-dependence and do not describe the data, while the data are well reproduced by the full EPOS 3 model calculations thereby showing the importance of the hadronic cascade phase at LHC energies. This observation agrees with the conclusion from the three-dimensional K$^\pm$ femtoscopic analysis in Pb--Pb collisions at $\sqrt{s_{\rm NN}}=2.76$~TeV \cite{Acharya:2017qtq} where the hydro-kinetic model HKM \cite{Shapoval:2014wya} with the hadronic rescattering phase described the charged kaon femtoscopic radii well.
The extracted experimental $\lambda$ values are about 0.45, whereas the EPOS 3 ones are about 0.65, i.e.\ apparently larger than the experimental $\lambda$ values.
The value of the $\lambda$ parameter may be influenced by non-Gaussian features of the
correlation function \cite{Morita:1999vj}, by a finite coherent component of kaon emission \cite{Abelev:2013pqa,Akkelin:2001nd} and also the contribution of kaons from K$^*$ decays ($\Gamma\sim$50~MeV, where $\Gamma$ is the decay width) and from other long-lived resonances \cite{Wiedemann:1996ig}.
The reason for the difference between the experimental correlation strengths and those obtained with EPOS 3 could be that the model does not accurately account for all contributions of kaons from various resonance decays \cite{Werner:2018yad}. Another explanation could be a partial coherence of the real emitting source \cite{Gyulassy:1979yi,Fowler:1977rq,Akkelin:2001nd,Bowler:1986ta}, which is not taken into account in the EPOS 3 model.

\begin{figure}[h]
\begin{center}
\includegraphics[width=0.5\textwidth]{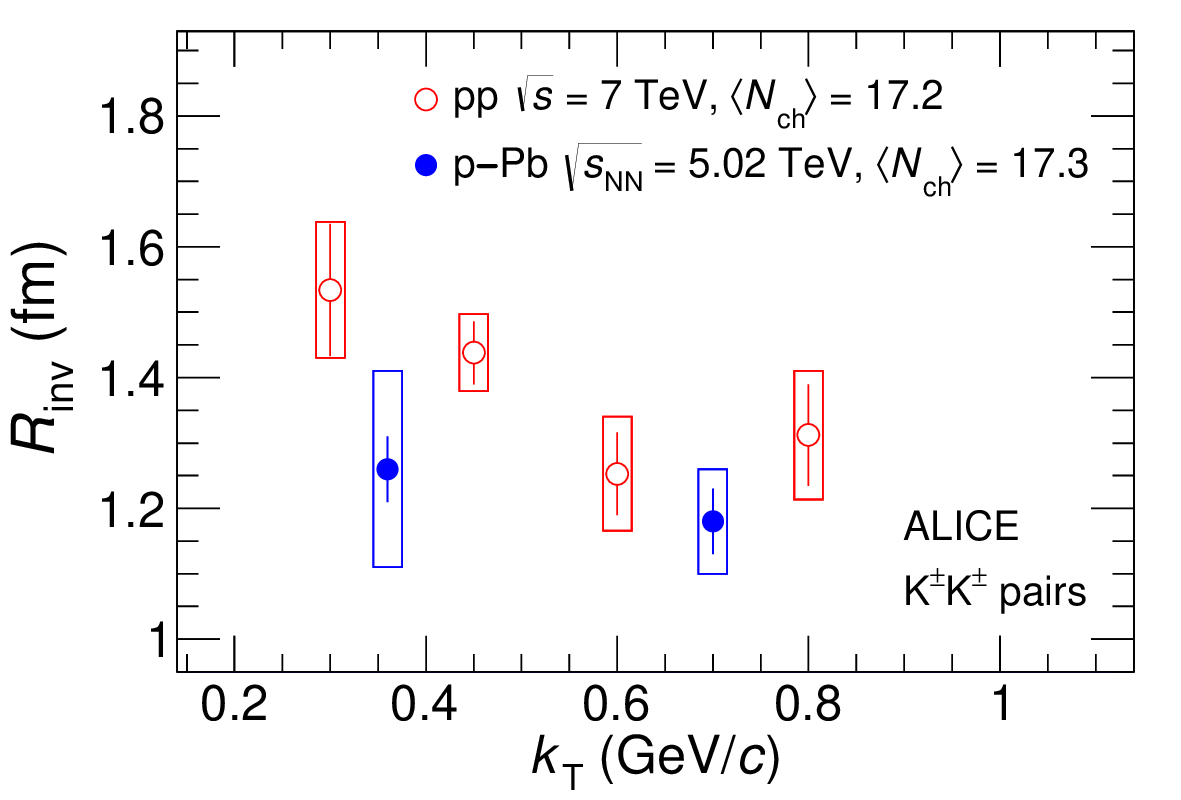}
\caption{Comparison of femtoscopic radii, as a function of pair transverse momentum $k_{\rm T}$,
obtained in pp \cite{Abelev:2012sq} and p--Pb collisions. Statistical (lines) and systematic uncertainties (boxes) are shown.}
\label{fig:Rinv_kt_ncheqv}
\end{center}
\end{figure}

In Fig.~\ref{fig:Rinv_kt_ncheqv}, the radii from pp collisions at $\sqrt{s}=7$~TeV \cite{Abelev:2012sq} and p--Pb collisions at $\sqrt{s_{\rm NN}}=5.02$~TeV at similar multiplicity are compared as a function of pair transverse momentum $k_{\rm T}$. The corresponding radii in Pb--Pb collisions at $\sqrt{s_{\rm NN}}=2.76$~TeV \cite{Adam:2015vja} are not shown
because they were obtained for multiplicities which are not available in this study.
The figure shows that at the same multiplicity, the radii in p--Pb collisions are consistent with those in pp collisions within uncertainties. The statistical significance of this observation (4--15\%) does not allow this result to be precisely compared with the results of the one-dimensional three-pion cumulants \cite{Abelev:2014pja} and three-dimensional two-pion \cite{Adam:2015pya} analyses where the radii in pp collisions were obtained to be 5--15\% and 10--20\% smaller than those in p--Pb collisions, respectively.

\begin{figure}
\centering
\subfigure{\label{fig:Rinv_NchA}\includegraphics[width=0.49\textwidth]{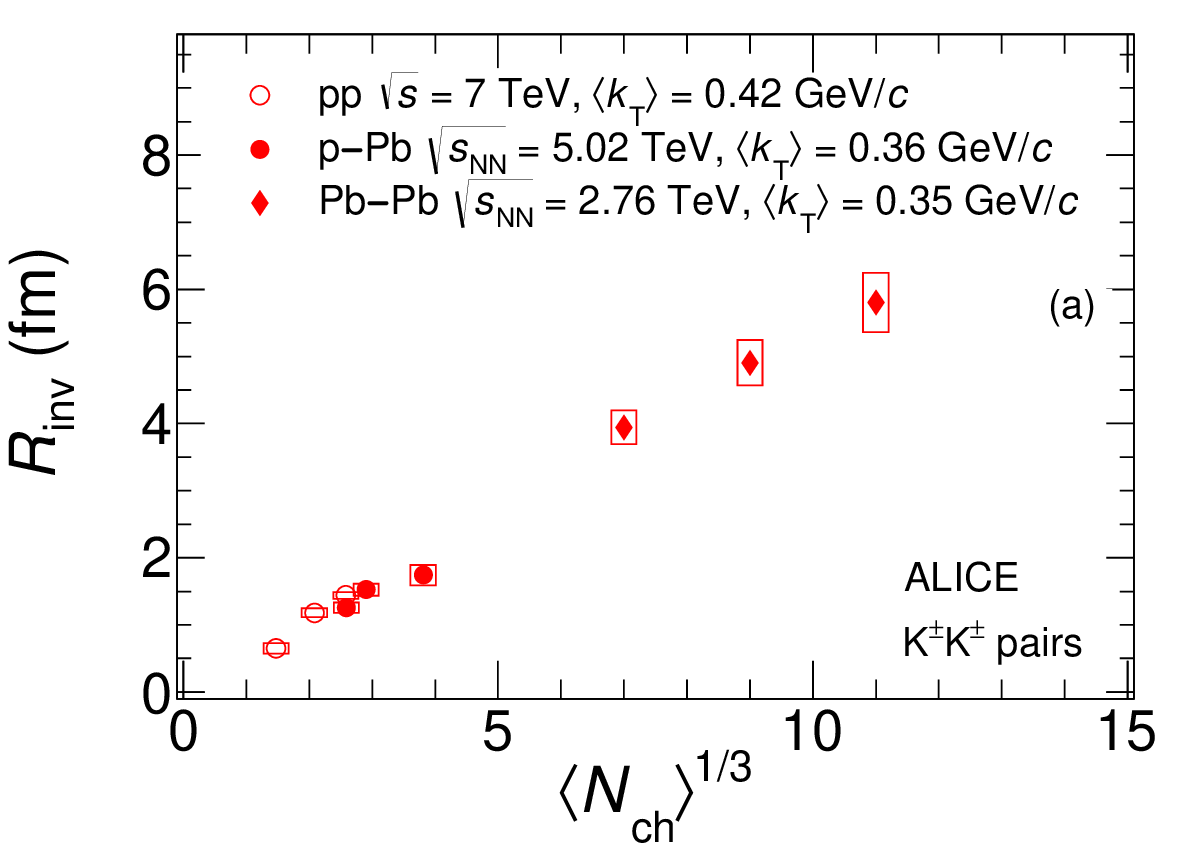}}~~~~~
\subfigure{\label{fig:Rinv_NchB}\includegraphics[width=0.49\textwidth]{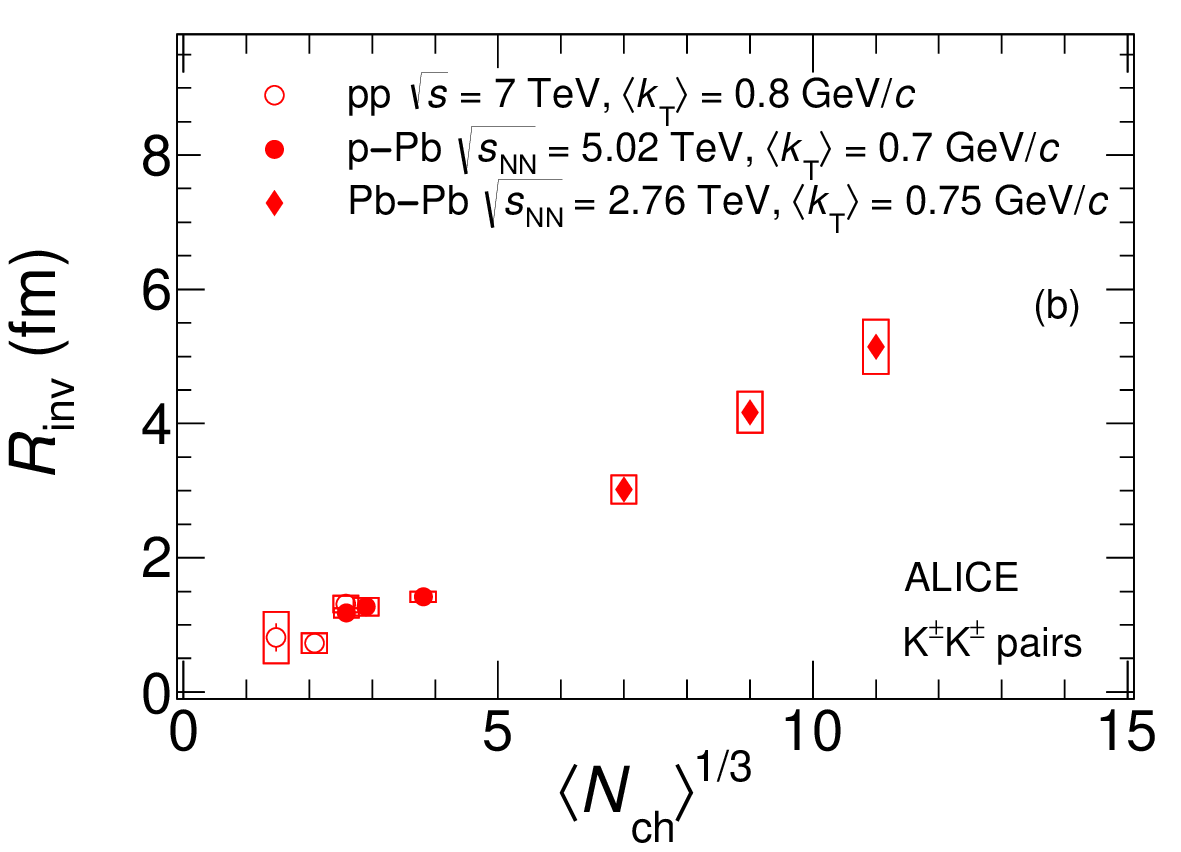}
}
\caption{Comparison of femtoscopic radii, as a function of the measured charged-particle multiplicity density $\langle N_{\rm ch}\rangle^{1/3}$, at low (a) and high (b) $k_{\rm T}$ obtained in pp \cite{Abelev:2012sq}, p--Pb and Pb--Pb \cite{Adam:2015vja} collisions. Statistical (lines) and systematic uncertainties (boxes) are shown.}
\label{fig:Rinv_Nch}
\end{figure}

Figure \ref{fig:Rinv_Nch} compares femtoscopic radii as a function of the measured charged-particle multiplicity density $\langle N_{\rm ch}\rangle^{1/3}$, at low (Fig.~\ref{fig:Rinv_NchA}) and high (Fig.~\ref{fig:Rinv_NchB}) $k_{\rm T}$ in pp \cite{Abelev:2012sq}, p--Pb and Pb--Pb \cite{Adam:2015vja} collisions. The obtained radii increase with $N_{\rm ch}$ and follow the multiplicity trend observed in pp collisions. The radii are equal in p--Pb and pp collisions at similar multiplicity within uncertainties.
This result could indicate that the dynamics of the source in p--Pb collisions at low multiplicities is similar to that in pp collisions. In particular, if there is a collective expansion of the sources created in pp and p--Pb collisions, these results indicate that the expansion is not significantly stronger in p--Pb than in pp collisions \cite{Bozek:2013df}.
As seen from the figure, the radii in p--Pb and Pb--Pb collisions were obtained in very different ranges of multiplicity and cannot be compared at the same $N_{\rm ch}$. In order to make a stronger conclusion between different collision systems, as was done in the pion correlation analyses \cite{Adam:2015pya,Abelev:2014pja}, a larger experimental data set should be considered.

\begin{figure}
\centering
\subfigure{\label{fig:Lam_NchA}\includegraphics[width=0.49\textwidth]{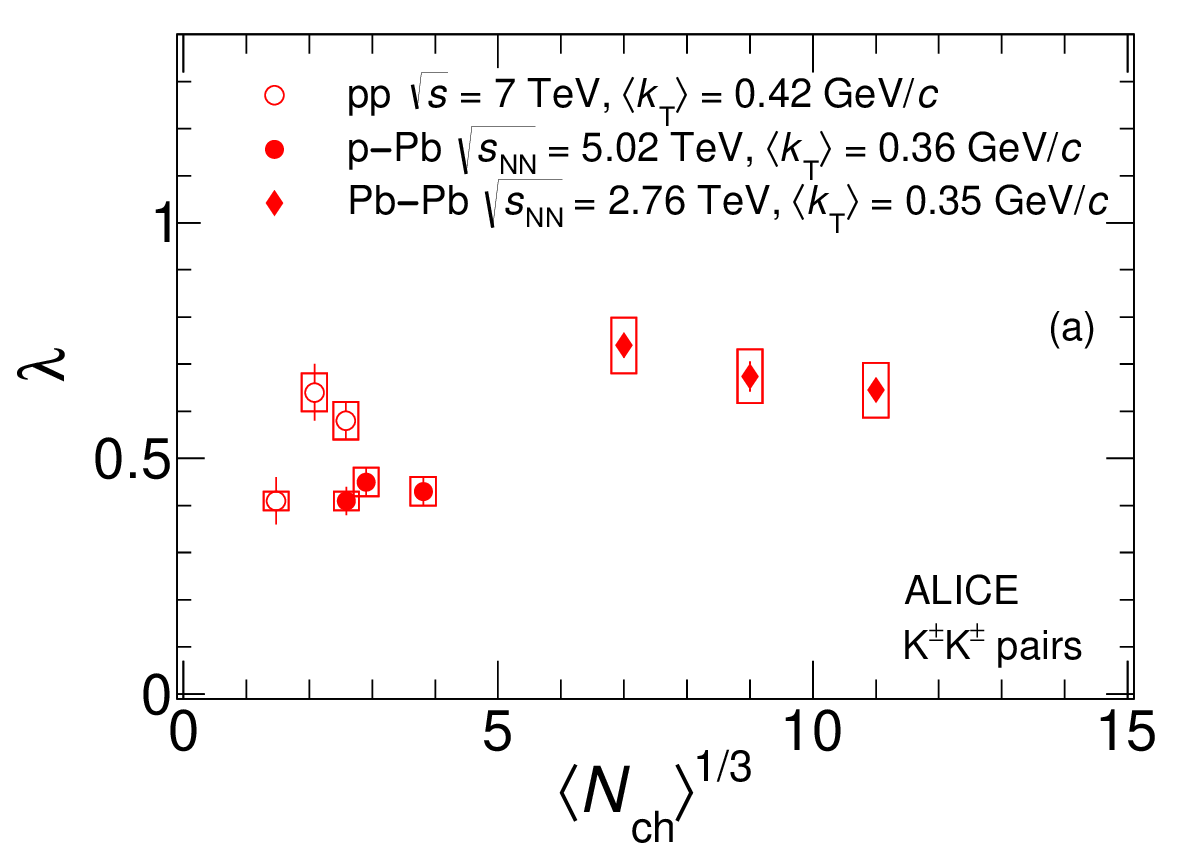}}
\subfigure{\label{fig:Lam_NchB}\includegraphics[width=0.49\textwidth]{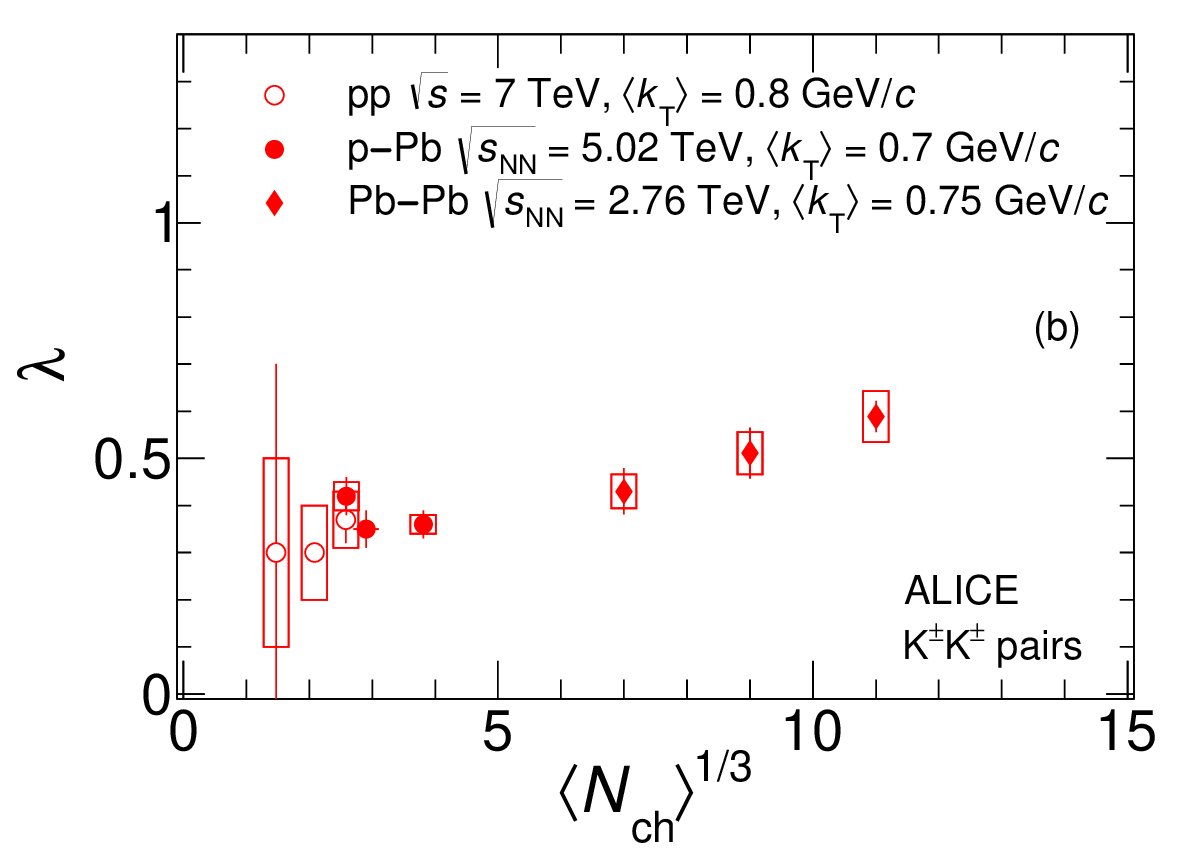}}
\caption{Comparison of correlation strengths $\lambda$, as a function of the measured charged-particle multiplicity density $\langle N_{\rm ch}\rangle^{1/3}$, at low (a) and high (b) $k_{\rm T}$
obtained in pp \cite{Abelev:2012sq}, p--Pb and Pb--Pb \cite{Adam:2015vja} collisions. Statistical (lines) and systematic uncertainties (boxes) are shown.}
\label{fig:Lam_Nch}
\end{figure}

Figures \ref{fig:Lam_NchA} and \ref{fig:Lam_NchB} show the correlation strengths $\lambda$ in pp \cite{Abelev:2012sq}, p--Pb and Pb--Pb \cite{Adam:2015vja} collisions at low and high $k_{\rm T}$, respectively.
All $\lambda$ values are less than unity
probably due to the influence of long-lived resonances and a non-Gaussian shape of the kaon CF peak. It can be noticed from the figure that the correlation strength parameters in Pb--Pb collisions tend to be higher than those in pp and p--Pb collisions. That could point to a more Gaussian source created in Pb--Pb collisions.

\begin{figure}[h]
\begin{center}
\includegraphics[width=0.7\textwidth]{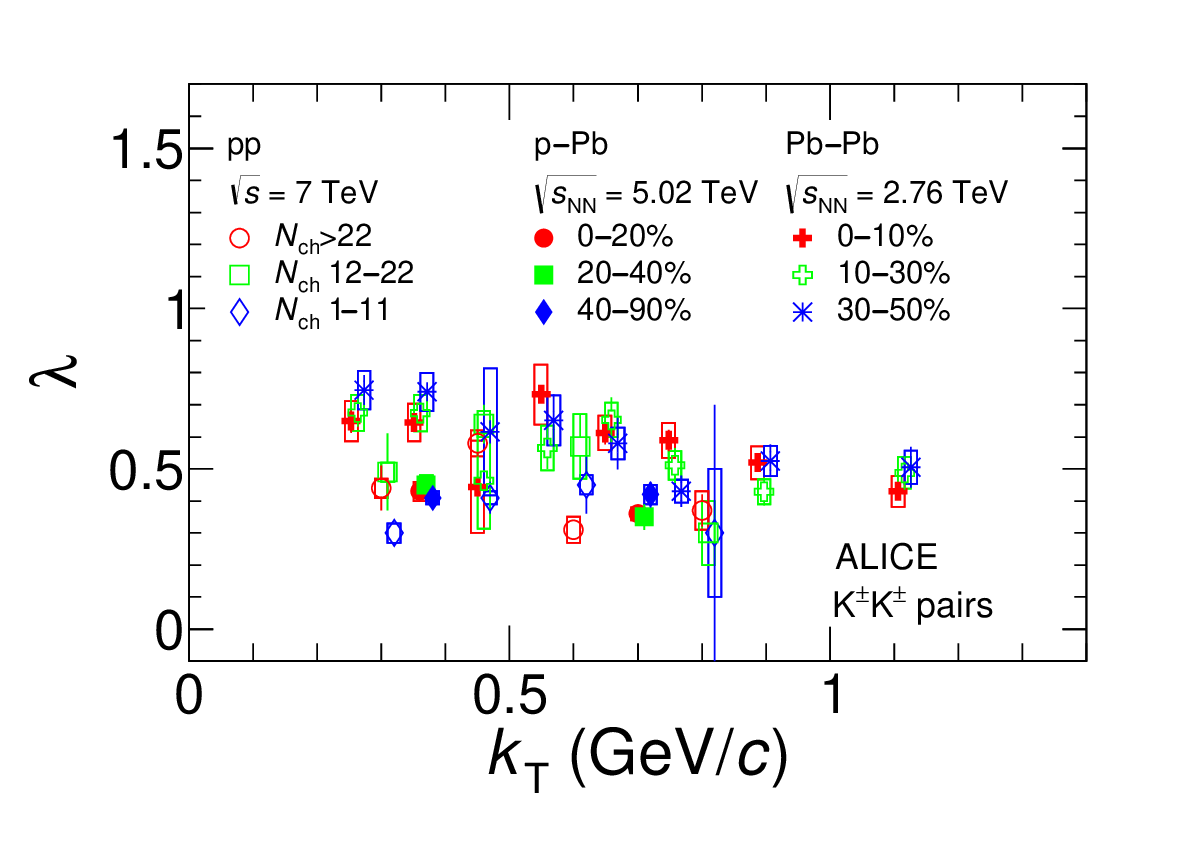}
\caption{(color online) The K$^{\pm}$K$^{\pm}$ correlation strengths $\lambda$ in pp \cite{Abelev:2012sq}, p--Pb and Pb--Pb \cite{Adam:2015vja} collisions versus pair transverse momentum $k_{\rm T}$ in all multiplicity and $k_{\rm T}$ bins. Statistical (lines) and systematic uncertainties (boxes) are shown. The data points for lower multiplicity classes (blue and green symbols) are slightly offset in $k_{\rm T}$ with respect to the highest multiplicity classes (red symbols) for better visibility.}
\label{fig:Lam_all}
\end{center}
\end{figure}

Figure \ref{fig:Lam_all} compares correlation strengths $\lambda$ in pp \cite{Abelev:2012sq}, p--Pb and Pb--Pb collisions as a function of $k_{\rm T}$ for all available multiplicity bins. As seen from the figure, the correlation strengths in all multiplicity and all $k_{\rm T}$ bins
do not show any noticeable $k_{\rm T}$ or multiplicity dependence. The systematic uncertainty values obtained for the compared collision systems are visibly different since even the same source of uncertainty gives a rather different contribution to the total uncertainty value in every collision system.

\section{Summary}
In this work, one-dimensional identical charged kaon correlations were obtained and analyzed for the first time in proton--nucleus collisions, that is in p--Pb at $\sqrt{s_{\rm NN}}=5.02$~TeV. The source size $R_{\rm inv}$ and correlation strength $\lambda$ were extracted from a correlation function parametrized in terms of the invariant pair relative momentum $q_{\mathrm{inv}}$. The obtained radii $R_{\rm inv}$ decrease with increasing pair transverse momentum $k_{\rm T}$ and with decreasing event multiplicity. This is similar to the behavior of pion radii in the three-dimensional two-pion correlation analysis in p--Pb collisions at $\sqrt{s_{\rm NN}}=5.02$~TeV
and one-dimensional three-pion cumulant results in pp collisions at $\sqrt{s}=7$~TeV, p--Pb collisions at $\sqrt{s_{\rm NN}}=5.02$~TeV and Pb--Pb collisions at $\sqrt{s_{\rm NN}}=2.76$~TeV.

The obtained radii $R_{\rm inv}$ are reproduced well by the EPOS 3 model (including femtoscopic effects) calculations with the hadronic rescattering phase, whose importance was also demonstrated in the three-dimensional femtoscopic analysis of K$^\pm$ pair correlations in Pb--Pb collisions. The values of the correlation strength parameters $\lambda$ in EPOS 3 are apparently larger than the experimental $\lambda$ values, which could be due to coherent sources not incorporated in EPOS 3 and long-lived resonances not taken into account accurately enough in this model.

The kaon $R_{\rm inv}$ values in p--Pb and pp collisions show the same trend with multiplicity. However, it is difficult to say whether the same is true for the Pb--Pb points because of a large gap in multiplicities available in p--Pb and Pb--Pb collisions. The results disfavor models which incorporate substantially stronger collective expansion in p--Pb collisions compared to pp collisions at similar multiplicity. The correlation strength $\lambda$ does not show any trends with
multiplicity or $k_{\rm T}$. The fact that the correlation strength in Pb--Pb collisions tends to be higher than in pp and p--Pb collisions could be an indication of a more Gaussian source created in Pb--Pb collisions. However, a stronger conclusion is prevented due to large statistical and systematic uncertainties, especially for the Pb--Pb data.

\newenvironment{acknowledgement}{\relax}{\relax}
\begin{acknowledgement}
\section*{Acknowledgements}
\input{fa_2019-03-04.tex}    
\end{acknowledgement}

\bibliography{biblio}

\newpage
\appendix
\section{The ALICE Collaboration}
\label{app:collab}
\input{2019-03-04-Alice_Authorlist_2019-Mar-04.tex}  
\end{document}

%% file: fa_2019-03-04.tex

The ALICE Collaboration would like to thank all its engineers and technicians for their invaluable contributions to the construction of the experiment and the CERN accelerator teams for the outstanding performance of the LHC complex.
The ALICE Collaboration gratefully acknowledges the resources and support provided by all Grid centres and the Worldwide LHC Computing Grid (WLCG) collaboration.
The ALICE Collaboration acknowledges the following funding agencies for their support in building and running the ALICE detector:
A. I. Alikhanyan National Science Laboratory (Yerevan Physics Institute) Foundation (ANSL), State Committee of Science and World Federation of Scientists (WFS), Armenia;
Austrian Academy of Sciences, Austrian Science Fund (FWF): [M 2467-N36] and Nationalstiftung f\"{u}r Forschung, Technologie und Entwicklung, Austria;
Ministry of Communications and High Technologies, National Nuclear Research Center, Azerbaijan;
Conselho Nacional de Desenvolvimento Cient\'{\i}fico e Tecnol\'{o}gico (CNPq), Universidade Federal do Rio Grande do Sul (UFRGS), Financiadora de Estudos e Projetos (Finep) and Funda\c{c}\~{a}o de Amparo \`{a} Pesquisa do Estado de S\~{a}o Paulo (FAPESP), Brazil;
Ministry of Science \& Technology of China (MSTC), National Natural Science Foundation of China (NSFC) and Ministry of Education of China (MOEC) , China;
Croatian Science Foundation and Ministry of Science and Education, Croatia;
Centro de Aplicaciones Tecnol\'{o}gicas y Desarrollo Nuclear (CEADEN), Cubaenerg\'{\i}a, Cuba;
Ministry of Education, Youth and Sports of the Czech Republic, Czech Republic;
The Danish Council for Independent Research | Natural Sciences, the Carlsberg Foundation and Danish National Research Foundation (DNRF), Denmark;
Helsinki Institute of Physics (HIP), Finland;
Commissariat \`{a} l'Energie Atomique (CEA), Institut National de Physique Nucl\'{e}aire et de Physique des Particules (IN2P3) and Centre National de la Recherche Scientifique (CNRS) and Rl\'{e}gion des  Pays de la Loire, France;
Bundesministerium f\"{u}r Bildung, Wissenschaft, Forschung und Technologie (BMBF) and GSI Helmholtzzentrum f\"{u}r Schwerionenforschung GmbH, Germany;
General Secretariat for Research and Technology, Ministry of Education, Research and Religions, Greece;
National Research, Development and Innovation Office, Hungary;
Department of Atomic Energy Government of India (DAE), Department of Science and Technology, Government of India (DST), University Grants Commission, Government of India (UGC) and Council of Scientific and Industrial Research (CSIR), India;
Indonesian Institute of Science, Indonesia;
Centro Fermi - Museo Storico della Fisica e Centro Studi e Ricerche Enrico Fermi and Istituto Nazionale di Fisica Nucleare (INFN), Italy;
Institute for Innovative Science and Technology , Nagasaki Institute of Applied Science (IIST), Japan Society for the Promotion of Science (JSPS) KAKENHI and Japanese Ministry of Education, Culture, Sports, Science and Technology (MEXT), Japan;
Consejo Nacional de Ciencia (CONACYT) y Tecnolog\'{i}a, through Fondo de Cooperaci\'{o}n Internacional en Ciencia y Tecnolog\'{i}a (FONCICYT) and Direcci\'{o}n General de Asuntos del Personal Academico (DGAPA), Mexico;
Nederlandse Organisatie voor Wetenschappelijk Onderzoek (NWO), Netherlands;
The Research Council of Norway, Norway;
Commission on Science and Technology for Sustainable Development in the South (COMSATS), Pakistan;
Pontificia Universidad Cat\'{o}lica del Per\'{u}, Peru;
Ministry of Science and Higher Education and National Science Centre, Poland;
Korea Institute of Science and Technology Information and National Research Foundation of Korea (NRF), Republic of Korea;
Ministry of Education and Scientific Research, Institute of Atomic Physics and Ministry of Research and Innovation and Institute of Atomic Physics, Romania;
Joint Institute for Nuclear Research (JINR), Ministry of Education and Science of the Russian Federation, National Research Centre Kurchatov Institute, Russian Science Foundation and Russian Foundation for Basic Research, Russia;
Ministry of Education, Science, Research and Sport of the Slovak Republic, Slovakia;
National Research Foundation of South Africa, South Africa;
Swedish Research Council (VR) and Knut \& Alice Wallenberg Foundation (KAW), Sweden;
European Organization for Nuclear Research, Switzerland;
National Science and Technology Development Agency (NSDTA), Suranaree University of Technology (SUT) and Office of the Higher Education Commission under NRU project of Thailand, Thailand;
Turkish Atomic Energy Agency (TAEK), Turkey;
National Academy of  Sciences of Ukraine, Ukraine;
Science and Technology Facilities Council (STFC), United Kingdom;
National Science Foundation of the United States of America (NSF) and United States Department of Energy, Office of Nuclear Physics (DOE NP), United States of America.

%% file: 2019-03-04-Alice_Authorlist_2019-Mar-04.tex

\begingroup
\small
\begin{flushleft}
S.~Acharya\Irefn{org141}\And 
D.~Adamov\'{a}\Irefn{org93}\And 
S.P.~Adhya\Irefn{org141}\And 
A.~Adler\Irefn{org74}\And 
J.~Adolfsson\Irefn{org80}\And 
M.M.~Aggarwal\Irefn{org98}\And 
G.~Aglieri Rinella\Irefn{org34}\And 
M.~Agnello\Irefn{org31}\And 
N.~Agrawal\Irefn{org10}\And 
Z.~Ahammed\Irefn{org141}
% ...
\And
\renewcommand\labelenumi{\textsuperscript{\theenumi}~}

\section*{Affiliation notes}
\renewcommand\theenumi{\roman{enumi}}
\begin{Authlist}
\item \Adef{org*}Deceased
\item \Adef{orgI}Dipartimento DET del Politecnico di Torino, Turin, Italy
\item \Adef{orgII}M.V. Lomonosov Moscow State University, D.V. Skobeltsyn Institute of Nuclear, Physics, Moscow, Russia
\item \Adef{orgIII}Department of Applied Physics, Aligarh Muslim University, Aligarh, India
\item \Adef{orgIV}Institute of Theoretical Physics, University of Wroclaw, Poland
\end{Authlist}

\section*{Collaboration Institutes}
\renewcommand\theenumi{\arabic{enumi}~}
\begin{Authlist}
\item \Idef{org1}A.I. Alikhanyan National Science Laboratory (Yerevan Physics Institute) Foundation, Yerevan, Armenia
\item \Idef{org2}Bogolyubov Institute for Theoretical Physics, National Academy of Sciences of Ukraine, Kiev, Ukraine
\item \Idef{org3}Bose Institute, Department of Physics  and Centre for Astroparticle Physics and Space Science (CAPSS), Kolkata, India
\item \Idef{org4}Budker Institute for Nuclear Physics, Novosibirsk, Russia
\item \Idef{org5}California Polytechnic State University, San Luis Obispo, California, United States
\item \Idef{org6}Central China Normal University, Wuhan, China
\item \Idef{org7}Centre de Calcul de l'IN2P3, Villeurbanne, Lyon, France
\item \Idef{org8}Centro de Aplicaciones Tecnol\'{o}gicas y Desarrollo Nuclear (CEADEN), Havana, Cuba
\item \Idef{org9}Centro de Investigaci\'{o}n y de Estudios Avanzados (CINVESTAV), Mexico City and M\'{e}rida, Mexico
\item \Idef{org10}Centro Fermi - Museo Storico della Fisica e Centro Studi e Ricerche ``Enrico Fermi', Rome, Italy
\item \Idef{org11}Chicago State University, Chicago, Illinois, United States
\item \Idef{org12}China Institute of Atomic Energy, Beijing, China
\item \Idef{org13}Chonbuk National University, Jeonju, Republic of Korea
\item \Idef{org14}Comenius University Bratislava, Faculty of Mathematics, Physics and Informatics, Bratislava, Slovakia
\item \Idef{org15}COMSATS University Islamabad, Islamabad, Pakistan
\item \Idef{org16}Creighton University, Omaha, Nebraska, United States
\item \Idef{org17}Department of Physics, Aligarh Muslim University, Aligarh, India
\item \Idef{org18}Department of Physics, Pusan National University, Pusan, Republic of Korea
\item \Idef{org19}Department of Physics, Sejong University, Seoul, Republic of Korea
\item \Idef{org20}Department of Physics, University of California, Berkeley, California, United States
\item \Idef{org21}Department of Physics, University of Oslo, Oslo, Norway
\item \Idef{org22}Department of Physics and Technology, University of Bergen, Bergen, Norway
\item \Idef{org23}Dipartimento di Fisica dell'Universit\`{a} 'La Sapienza' and Sezione INFN, Rome, Italy
\item \Idef{org24}Dipartimento di Fisica dell'Universit\`{a} and Sezione INFN, Cagliari, Italy
\item \Idef{org25}Dipartimento di Fisica dell'Universit\`{a} and Sezione INFN, Trieste, Italy
\item \Idef{org26}Dipartimento di Fisica dell'Universit\`{a} and Sezione INFN, Turin, Italy
\item \Idef{org27}Dipartimento di Fisica e Astronomia dell'Universit\`{a} and Sezione INFN, Bologna, Italy
\item \Idef{org28}Dipartimento di Fisica e Astronomia dell'Universit\`{a} and Sezione INFN, Catania, Italy
\item \Idef{org29}Dipartimento di Fisica e Astronomia dell'Universit\`{a} and Sezione INFN, Padova, Italy
\item \Idef{org30}Dipartimento di Fisica `E.R.~Caianiello' dell'Universit\`{a} and Gruppo Collegato INFN, Salerno, Italy
\item \Idef{org31}Dipartimento DISAT del Politecnico and Sezione INFN, Turin, Italy
\item \Idef{org32}Dipartimento di Scienze e Innovazione Tecnologica dell'Universit\`{a} del Piemonte Orientale and INFN Sezione di Torino, Alessandria, Italy
\item \Idef{org33}Dipartimento Interateneo di Fisica `M.~Merlin' and Sezione INFN, Bari, Italy
\item \Idef{org34}European Organization for Nuclear Research (CERN), Geneva, Switzerland
\item \Idef{org35}Faculty of Electrical Engineering, Mechanical Engineering and Naval Architecture, University of Split, Split, Croatia
\item \Idef{org36}Faculty of Engineering and Science, Western Norway University of Applied Sciences, Bergen, Norway
\item \Idef{org37}Faculty of Nuclear Sciences and Physical Engineering, Czech Technical University in Prague, Prague, Czech Republic
\item \Idef{org38}Faculty of Science, P.J.~\v{S}af\'{a}rik University, Ko\v{s}ice, Slovakia
\item \Idef{org39}Frankfurt Institute for Advanced Studies, Johann Wolfgang Goethe-Universit\"{a}t Frankfurt, Frankfurt, Germany
\item \Idef{org40}Gangneung-Wonju National University, Gangneung, Republic of Korea
\item \Idef{org41}Gauhati University, Department of Physics, Guwahati, India
\item \Idef{org42}Helmholtz-Institut f\"{u}r Strahlen- und Kernphysik, Rheinische Friedrich-Wilhelms-Universit\"{a}t Bonn, Bonn, Germany
\item \Idef{org43}Helsinki Institute of Physics (HIP), Helsinki, Finland
\item \Idef{org44}High Energy Physics Group,  Universidad Aut\'{o}noma de Puebla, Puebla, Mexico
\item \Idef{org45}Hiroshima University, Hiroshima, Japan
\item \Idef{org46}Hochschule Worms, Zentrum  f\"{u}r Technologietransfer und Telekommunikation (ZTT), Worms, Germany
\item \Idef{org47}Horia Hulubei National Institute of Physics and Nuclear Engineering, Bucharest, Romania
\item \Idef{org48}Indian Institute of Technology Bombay (IIT), Mumbai, India
\item \Idef{org49}Indian Institute of Technology Indore, Indore, India
\item \Idef{org50}Indonesian Institute of Sciences, Jakarta, Indonesia
\item \Idef{org51}INFN, Laboratori Nazionali di Frascati, Frascati, Italy
\item \Idef{org52}INFN, Sezione di Bari, Bari, Italy
\item \Idef{org53}INFN, Sezione di Bologna, Bologna, Italy
\item \Idef{org54}INFN, Sezione di Cagliari, Cagliari, Italy
\item \Idef{org55}INFN, Sezione di Catania, Catania, Italy
\item \Idef{org56}INFN, Sezione di Padova, Padova, Italy
\item \Idef{org57}INFN, Sezione di Roma, Rome, Italy
\item \Idef{org58}INFN, Sezione di Torino, Turin, Italy
\item \Idef{org59}INFN, Sezione di Trieste, Trieste, Italy
\item \Idef{org60}Inha University, Incheon, Republic of Korea
\item \Idef{org61}Institut de Physique Nucl\'{e}aire d'Orsay (IPNO), Institut National de Physique Nucl\'{e}aire et de Physique des Particules (IN2P3/CNRS), Universit\'{e} de Paris-Sud, Universit\'{e} Paris-Saclay, Orsay, France
\item \Idef{org62}Institute for Nuclear Research, Academy of Sciences, Moscow, Russia
\item \Idef{org63}Institute for Subatomic Physics, Utrecht University/Nikhef, Utrecht, Netherlands
\item \Idef{org64}Institute for Theoretical and Experimental Physics, Moscow, Russia
\item \Idef{org65}Institute of Experimental Physics, Slovak Academy of Sciences, Ko\v{s}ice, Slovakia
\item \Idef{org66}Institute of Physics, Homi Bhabha National Institute, Bhubaneswar, India
\item \Idef{org67}Institute of Physics of the Czech Academy of Sciences, Prague, Czech Republic
\item \Idef{org68}Institute of Space Science (ISS), Bucharest, Romania
\item \Idef{org69}Institut f\"{u}r Kernphysik, Johann Wolfgang Goethe-Universit\"{a}t Frankfurt, Frankfurt, Germany
\item \Idef{org70}Instituto de Ciencias Nucleares, Universidad Nacional Aut\'{o}noma de M\'{e}xico, Mexico City, Mexico
\item \Idef{org71}Instituto de F\'{i}sica, Universidade Federal do Rio Grande do Sul (UFRGS), Porto Alegre, Brazil
\item \Idef{org72}Instituto de F\'{\i}sica, Universidad Nacional Aut\'{o}noma de M\'{e}xico, Mexico City, Mexico
\item \Idef{org73}iThemba LABS, National Research Foundation, Somerset West, South Africa
\item \Idef{org74}Johann-Wolfgang-Goethe Universit\"{a}t Frankfurt Institut f\"{u}r Informatik, Fachbereich Informatik und Mathematik, Frankfurt, Germany
\item \Idef{org75}Joint Institute for Nuclear Research (JINR), Dubna, Russia
\item \Idef{org76}Korea Institute of Science and Technology Information, Daejeon, Republic of Korea
\item \Idef{org77}KTO Karatay University, Konya, Turkey
\item \Idef{org78}Laboratoire de Physique Subatomique et de Cosmologie, Universit\'{e} Grenoble-Alpes, CNRS-IN2P3, Grenoble, France
\item \Idef{org79}Lawrence Berkeley National Laboratory, Berkeley, California, United States
\item \Idef{org80}Lund University Department of Physics, Division of Particle Physics, Lund, Sweden
\item \Idef{org81}Nagasaki Institute of Applied Science, Nagasaki, Japan
\item \Idef{org82}Nara Women{'}s University (NWU), Nara, Japan
\item \Idef{org83}National and Kapodistrian University of Athens, School of Science, Department of Physics , Athens, Greece
\item \Idef{org84}National Centre for Nuclear Research, Warsaw, Poland
\item \Idef{org85}National Institute of Science Education and Research, Homi Bhabha National Institute, Jatni, India
\item \Idef{org86}National Nuclear Research Center, Baku, Azerbaijan
\item \Idef{org87}National Research Centre Kurchatov Institute, Moscow, Russia
\item \Idef{org88}Niels Bohr Institute, University of Copenhagen, Copenhagen, Denmark
\item \Idef{org89}Nikhef, National institute for subatomic physics, Amsterdam, Netherlands
\item \Idef{org90}NRC Kurchatov Institute IHEP, Protvino, Russia
\item \Idef{org91}NRNU Moscow Engineering Physics Institute, Moscow, Russia
\item \Idef{org92}Nuclear Physics Group, STFC Daresbury Laboratory, Daresbury, United Kingdom
\item \Idef{org93}Nuclear Physics Institute of the Czech Academy of Sciences, \v{R}e\v{z} u Prahy, Czech Republic
\item \Idef{org94}Oak Ridge National Laboratory, Oak Ridge, Tennessee, United States
\item \Idef{org95}Ohio State University, Columbus, Ohio, United States
\item \Idef{org96}Petersburg Nuclear Physics Institute, Gatchina, Russia
\item \Idef{org97}Physics department, Faculty of science, University of Zagreb, Zagreb, Croatia
\item \Idef{org98}Physics Department, Panjab University, Chandigarh, India
\item \Idef{org99}Physics Department, University of Jammu, Jammu, India
\item \Idef{org100}Physics Department, University of Rajasthan, Jaipur, India
\item \Idef{org101}Physikalisches Institut, Eberhard-Karls-Universit\"{a}t T\"{u}bingen, T\"{u}bingen, Germany
\item \Idef{org102}Physikalisches Institut, Ruprecht-Karls-Universit\"{a}t Heidelberg, Heidelberg, Germany
\item \Idef{org103}Physik Department, Technische Universit\"{a}t M\"{u}nchen, Munich, Germany
\item \Idef{org104}Politecnico di Bari, Bari, Italy
\item \Idef{org105}Research Division and ExtreMe Matter Institute EMMI, GSI Helmholtzzentrum f\"ur Schwerionenforschung GmbH, Darmstadt, Germany
\item \Idef{org106}Rudjer Bo\v{s}kovi\'{c} Institute, Zagreb, Croatia
\item \Idef{org107}Russian Federal Nuclear Center (VNIIEF), Sarov, Russia
\item \Idef{org108}Saha Institute of Nuclear Physics, Homi Bhabha National Institute, Kolkata, India
\item \Idef{org109}School of Physics and Astronomy, University of Birmingham, Birmingham, United Kingdom
\item \Idef{org110}Secci\'{o}n F\'{\i}sica, Departamento de Ciencias, Pontificia Universidad Cat\'{o}lica del Per\'{u}, Lima, Peru
\item \Idef{org111}Shanghai Institute of Applied Physics, Shanghai, China
\item \Idef{org112}St. Petersburg State University, St. Petersburg, Russia
\item \Idef{org113}Stefan Meyer Institut f\"{u}r Subatomare Physik (SMI), Vienna, Austria
\item \Idef{org114}SUBATECH, IMT Atlantique, Universit\'{e} de Nantes, CNRS-IN2P3, Nantes, France
\item \Idef{org115}Suranaree University of Technology, Nakhon Ratchasima, Thailand
\item \Idef{org116}Technical University of Ko\v{s}ice, Ko\v{s}ice, Slovakia
\item \Idef{org117}Technische Universit\"{a}t M\"{u}nchen, Excellence Cluster 'Universe', Munich, Germany
\item \Idef{org118}The Henryk Niewodniczanski Institute of Nuclear Physics, Polish Academy of Sciences, Cracow, Poland
\item \Idef{org119}The University of Texas at Austin, Austin, Texas, United States
\item \Idef{org120}Universidad Aut\'{o}noma de Sinaloa, Culiac\'{a}n, Mexico
\item \Idef{org121}Universidade de S\~{a}o Paulo (USP), S\~{a}o Paulo, Brazil
\item \Idef{org122}Universidade Estadual de Campinas (UNICAMP), Campinas, Brazil
\item \Idef{org123}Universidade Federal do ABC, Santo Andre, Brazil
\item \Idef{org124}University College of Southeast Norway, Tonsberg, Norway
\item \Idef{org125}University of Cape Town, Cape Town, South Africa
\item \Idef{org126}University of Houston, Houston, Texas, United States
\item \Idef{org127}University of Jyv\"{a}skyl\"{a}, Jyv\"{a}skyl\"{a}, Finland
\item \Idef{org128}University of Liverpool, Liverpool, United Kingdom
\item \Idef{org129}University of Science and Techonology of China, Hefei, China
\item \Idef{org130}University of Tennessee, Knoxville, Tennessee, United States
\item \Idef{org131}University of the Witwatersrand, Johannesburg, South Africa
\item \Idef{org132}University of Tokyo, Tokyo, Japan
\item \Idef{org133}University of Tsukuba, Tsukuba, Japan
\item \Idef{org134}Universit\'{e} Clermont Auvergne, CNRS/IN2P3, LPC, Clermont-Ferrand, France
\item \Idef{org135}Universit\'{e} de Lyon, Universit\'{e} Lyon 1, CNRS/IN2P3, IPN-Lyon, Villeurbanne, Lyon, France
\item \Idef{org136}Universit\'{e} de Strasbourg, CNRS, IPHC UMR 7178, F-67000 Strasbourg, France, Strasbourg, France
\item \Idef{org137}Universit\'{e} Paris-Saclay Centre d'Etudes de Saclay (CEA), IRFU, D\'{e}partment de Physique Nucl\'{e}aire (DPhN), Saclay, France
\item \Idef{org138}Universit\`{a} degli Studi di Foggia, Foggia, Italy
\item \Idef{org139}Universit\`{a} degli Studi di Pavia, Pavia, Italy
\item \Idef{org140}Universit\`{a} di Brescia, Brescia, Italy
\item \Idef{org141}Variable Energy Cyclotron Centre, Homi Bhabha National Institute, Kolkata, India
\item \Idef{org142}Warsaw University of Technology, Warsaw, Poland
\item \Idef{org143}Wayne State University, Detroit, Michigan, United States
\item \Idef{org144}Westf\"{a}lische Wilhelms-Universit\"{a}t M\"{u}nster, Institut f\"{u}r Kernphysik, M\"{u}nster, Germany
\item \Idef{org145}Wigner Research Centre for Physics, Hungarian Academy of Sciences, Budapest, Hungary
\item \Idef{org146}Yale University, New Haven, Connecticut, United States
\item \Idef{org147}Yonsei University, Seoul, Republic of Korea
\end{Authlist}
\endgroup